\documentclass[aps,prl,superscriptaddress,floatfix,twocolumn,showpacs]{revtex4-1}

\pdfoutput=1

\usepackage{amsmath,dcolumn,bm,amsthm,amsfonts,amssymb,tabularx,subfigure}
\usepackage[colorlinks=true,urlcolor=blue,citecolor=blue,linkcolor=blue]{hyperref}
\usepackage{mathrsfs,times}
\usepackage{bbold,dsfont}
\usepackage{graphicx,graphics,color}
\usepackage{mathtools,nicefrac}
\usepackage{slashed}
\allowdisplaybreaks

\newcommand{\beq}{\begin{equation}}
\newcommand{\eeq}{\end{equation}}
\newcommand{\bea}{\begin{eqnarray}}
\newcommand{\eea}{\end{eqnarray}}
\newcommand{\widebar}{\overline}

\newcommand{\R}{\textrm{R}}
\newcommand{\NS}{\textrm{NS}}
\newcommand{\XZ}{{X\!Z}}
\newcommand\ddfrac[2]{\nicefrac{\displaystyle #1}{\displaystyle #2}}

\newcommand{\fb}{{\mathrm{fb}}}

\begin{document}

\title{From supersymmetric sine-Gordon equation to the superconformal minimal model}

\author{Chun Chen}
\email[]{chun6@ualberta.ca}
\affiliation{Department of Physics, University of Alberta, Edmonton, Alberta T6G 2E1, Canada}

\date{\today}

\begin{abstract}

We propose a generalization of the Grover-Sheng-Vishwanath model which, as solved by the density-matrix renormalization group, realizes the emergent supersymmetric criticality in the universality class of the \emph{even} series of the $\mathcal{N}\!=\!1$ superconformal minimal models characterized by a central charge $5/4$. This chain model describes the topological phase transition of the propagating Majorana edge mode in topological superconductors coupled with the \emph{two}-flavour Ising magnetic fluctuations (or the $\XZ$ type). Using bosonization and perturbative renormalization group, we show that the augmented degrees of freedom trigger a paradigm shift from the supersymmetric Landau-Ginzburg action to the variant of the supersymmetric sine-Gordon equation, which, in the massless case, can flow towards the supersymmetric minimal series upon a generalized Feigin-Fuchs construction. Therefore, the present lattice model comprises a concrete system that exhibits the spacetime supersymmetry through the distinct route.    

\end{abstract}

\pacs{}

\maketitle

{\it Introduction.---}Supersymmetry (SUSY) is a beautiful invention and a powerful tool in theoretical physics \cite{WESS1974}. It extends the limit of geometric symmetries beyond the conventional spacetime and relates bosons and fermions through the extra Grassmannian dimensions \cite{Shifman}. Being flourished in gauge dynamics and high-energy physics, a persisting but outstanding question remains how to realize SUSY in nature. Instead of pursuing as the exact global symmetry of the universe, in the vicinity of continuous phase transitions, there exist known examples for which spacetime SUSY emerges at the critical point. This more statistical mechanical scenario receives renewed attention in the advent of topological materials, particularly those exotic critical phenomena pertaining to the transitions involving the anomalous boundary degrees of freedom. Exploiting this line of reasoning, a handful of microscopic models have been proposed to harbor the superconformal criticality \cite{BPZ,FQSPRL} in the tricritical Ising (TCI) universality class \cite{Feiguin,gils2009,GroverShengVishwanath,RahmaniPRL,Li,OBrienFendley,SannomiyaKatsura}. In $(1+1)$D, the simplest supersymmetric field theories belong to the category of the $\mathcal{N}\!=\!1$ superconformal minimal models (SMMs) \cite{FQS,BKT1985} whose central charges are given by
\beq
c(m)=\frac{3}{2}\left[1-\frac{8}{m(m+2)}\right],\ \ \ m=2,3,4,\ldots.
\eeq
The TCI model has central charge $7/10$, providing the first nontrivial representation of the SMM with $m\!=\!3$. Compared to the well-understood odd-$m$ series of SMMs, largely due to the availability of the flexible lattice TCI models, the even-$m$ series of SMMs are much less explored \cite{FriedanShenkerGaussian,yang1987,gils2009}.

In this Letter, we advance this pursuit by constructing such a microscopic model on a chain lattice to realize the $m\!=\!6$ SMM featuring a critical central charge $5/4$. This model generalizes the Grover-Sheng-Vishwanath (GSV) proposal \cite{GroverShengVishwanath} in considering the induced quantum phase transition from the interplay between the multi-flavour magnetic fluctuation and the gapless Majorana mode along the edge of $2$D topological superconductor. Utilizing bosonization and renormalization group (RG), we corroborate the density-matrix-renormalization-group (DMRG) observation on the endowed SUSY through establishing the connection between its continuum description and the supersymmetric sine-Gordon (SSG) model \cite{WITTEN1978} within a Coulomb-gas formalism \cite{DotsenkoFateev,BigYellow}. 

{\it Lattice model.---}The fermion-boson hybrid system may be described by the coupled chain Hamiltonian,
\begin{align}
H&=H_1+H_2+H_3, \label{hami} \\[0.5em]
H_1\!&=-\sum_{i}\!\left(\sigma^z_i\sigma^z_{i+1}+\sigma^x_i\right)\!,  \label{tfihami} \\
H_2\!&=-\sum_{i}\!\left[J_{zz}\!\left(u^z_{i,a}u^z_{i,b}+u^z_{i,b}u^z_{i+1,a}\right) \right. \nonumber \\
&\ \ \ \ \ \ \ \ \ \ \ \ \ \ \ \ \left.+J_{xx}\!\left(u^x_{i,a}u^x_{i,b}+u^x_{i,b}u^x_{i+1,a}\right)\right]\!,  \label{xzhami} \\[0.5em]
H_3\!&=\!\sum_{i}\!\left[g\sigma^x_i\!\left(u^z_{i,a}+u^z_{i,b}\right)\!-\!g\sigma^z_i\sigma^z_{i+1}\!\left(u^z_{i,b}+u^z_{i+1,a}\right)\right]\!,
\end{align}
where $\widehat{\boldsymbol{\sigma}}_i,\ \widehat{\boldsymbol{u}}_{i,a},\ \widehat{\boldsymbol{u}}_{i,b}$ are the commuting spin-$1/2$ Pauli matrices on site $i$. Via the Jordan-Wigner (JW) transformation, the massless Majorana hopping term $-i\sum_j\chi_j\chi_{j+1}$ can be mapped to a critical (i.e., self-dual) Ising chain $H_1$ of the $\widehat{\boldsymbol{\sigma}}$-spins subject to the transverse magnetic field. The original GSV's transverse-field-Ising magnetic fluctuation has been replaced by the $\XZ$-type spin-spin interaction in $H_2$. As will be explained below, this extension fundamentally alters the underlying physics of the model, particularly giving rise to the $\mathcal{N}\!=\!1$ supersymmetry of a different kind at criticality. The interplay between the propagating Majorana fermions and the fluctuating Ising spins retains a standard form of minimal coupling as is given by $H_3$.

\begin{figure*}[t]
\centering
\includegraphics[width=0.9\textwidth]{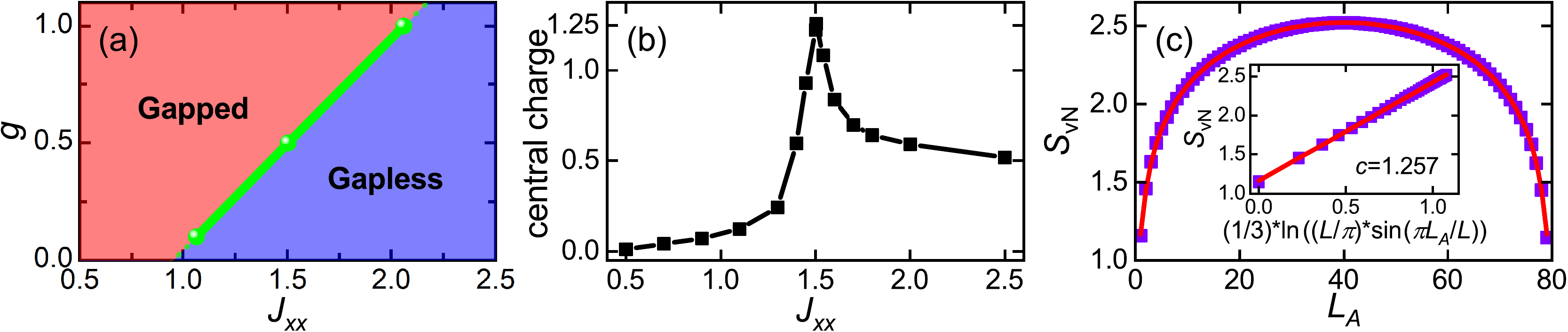}
\caption{\label{fig:fig1} DMRG results: (a) is the numerically calculated ground-state phase diagram of model (\ref{hami}) in the $J_{xx}$-$g$ plane with fixed $J_{zz}\!=\!1$. The blue gapless region $(c\!=\!1/2)$ is separated from the red gapped phase $(c\!=\!0)$ by a continuous transition marked by the green line where the SUSY criticality $(c\!=\!5/4)$ emerges. Panel (b) depicts the change of the central charge along the line cut $g\!=\!0.5$ in (a) as tuning $J_{xx}$ across the critical point $J^c_{xx}=1.504$. The low-$J_{xx}$ gapped and high-$J_{xx}$ gapless regions as well as the transition point featuring the maximal central charge $5/4$ can be identified therein. The same type of transition is reproduced along $g\!=\!0.1\ \mbox{and}\ 1$ as well. (c) illustrates the entanglement entropy $S_{\textrm{vN}}$ as a function of the segment length $L_A$ at $g\!=\!0.5$ and $J^c_{xx}=1.504$, from which the critical central charge is fitted to be $1.257$. The system size is up to $L\!=\!80$.} 
\end{figure*}

\begin{table*}[tb]
\setlength{\tabcolsep}{10pt}
\begin{tabular}{ c||ccccccccc  }
 \hline\hline
 \\[-0.7em]
 $\R$ & $\ddfrac{5}{16}$ & $\ddfrac{41}{96}$ & $\ddfrac{5}{96}$ & $\ddfrac{29}{16}$ & $\ddfrac{9}{16}$ & $\ddfrac{1}{16}$ & $\ddfrac{67}{32}$ & $\ddfrac{23}{32}$ & $\ddfrac{3}{32}$ \\ [0.5em]
 $\NS$ & $0$ & $\ddfrac{1}{32}$ & $\ddfrac{5}{6}$ & $\ddfrac{1}{12}$ & $\ddfrac{33}{32}$ & $\ddfrac{5}{32}$ & $3$ & $\ddfrac{5}{4}$ & $\ddfrac{1}{4}$ \\[0.3em]
 \hline \hline
\end{tabular}
  \caption{Operator content of the $m\!=\!6\ (c\!=\!5/4)$ SMM with periodic boundary conditions.}
    \label{tab:1}
\end{table*}

In terms of the dual spin variables (after relabeling $\widehat{\boldsymbol{u}}_{i,a}\rightarrow\widehat{\boldsymbol{u}}_{2i-1}$ and $\widehat{\boldsymbol{u}}_{i,b}\rightarrow\widehat{\boldsymbol{u}}_{2i}$),
\beq
\tau^x_{j+1}\coloneqq u^z_j u^z_{j+1},\ \ \ \tau^x_{j+1}\coloneqq\prod_{i<j+1} u^x_i, \label{dualspin}
\eeq
up to inessential boundary corrections, the $\XZ$ spin chain $H_2$ is equivalent to two noninteracting quantum Ising spin chains positioned alternatively across the even and odd lattice sites,
\beq
H_2=-\sum_{j} \left(J_{xx}\tau^z_{j}\tau^z_{j+2}+J_{zz}\tau^x_j\right).
\eeq
In this sense, the $\XZ$-type magnetic interaction we consider is dual to a two-flavour transverse-field-Ising fluctuation. Nevertheless, in view of the fact that the coupling term $H_3$ transforms nonlocally in the disorder operators $\widehat{\boldsymbol{\tau}}_j$, the nature of the transition is changed qualitatively.

For instance, suppose $g\!=\!0$, then because $H_1$ corresponds to the free real-fermion action with central charge $1/2$ and $H_2$ once fine-tuned to the $J_{zz}\!=\!J_{xx}$ point describes a free real scalar field carrying central charge $1$, it is conceivable from Zamolodchikov's $c$-theorem that for a probable second-order phase transition of $H$ occurring at finite $g$, the accompanying scaling invariance and unitarity constraint may facilitate the emergence of SUSY, which taken together dictate the realization of the $\mathcal{N}\!=\!1$ SMM at the critical point. Relative to the GSV model, the additional flavour in the $\XZ$ spin fluctuation might thus enable us to go beyond the SUSY of the TCI universality class.

{\it DMRG results.---}We solve for the ground state of Hamiltonian (\ref{hami}) and investigate the critical phenomena associated with the continuous phase transition thereof by the numerical DMRG method \cite{white1992}. For moderate chains $(L\!=\!30\!\sim\!60)$ the periodic boundary conditions (PBCs) can be imposed straightforwardly. To achieve higher accuracy, the technique of sine-square deformation \cite{SSD,Ishibashissd} desired for OBCs may be implemented when simulating the larger system's sizes of $L\geqslant80$.

Figure~\ref{fig:fig1}(a) shows the ground-state phase diagram of (\ref{hami}) on the $J_{xx}$-$g$ plane at constant $J_{zz}\!=\!1$ obtained from DMRG. The gapped phase (red region) with $c\!=\!0$ is stabilized provided the perturbing field $J_{xx}/J_{zz}$ is small. Then, due to the finite $g$-coupling, the Majorana mode is subsequently gapped out through acquiring a dynamically-generated mass term proportional to $\langle u^z_{i,a(b)}\rangle\!\neq\!0$. The situation is reversed however in the $J_{xx}$-dominated region, where the $u^z_{i,a(b)}$-spins become strongly disordered, rendering the fermion-boson coupling irrelevant. As a result, the effectively decoupled Majorana chain retrieves its criticality $(c\!=\!1/2)$ and the resulting gaplessness maintains thereafter under the protection of an antiunitary symmetry. Interestingly, more exotic phenomena can arise at the inbetween stage once the $\XZ$ chain (\ref{xzhami}) is tuned close to its criticality as well. The coherent resonance between the Ising and $\XZ$ criticality compromises exactly at the continuous transition point whose underlying CFT enhanced by the probable emergent SUSY must thus fall into the discrete list of the $\mathcal{N}\!=\!1$ SMMs of $1/2\!<\!c\!<\!3/2$. Indeed, in Fig.~\ref{fig:fig1}(b) the DMRG simulation on the evolution of the central charge along the horizontal line of $g\!=\!0.5$ in panel (a) confirms the existence of the critical region [denoted by the green line in (a)] where the maximal central charge reaches $1.257$ at the transition point $J^c_{xx}\!=\!1.504$ after fitting to the Calabrese-Cardy formula of the entanglement entropy \cite{CalabreseCardy} [see Fig.~\ref{fig:fig1}(c) and the inset],
\beq
S_{\textrm{vN}}(L_A)=\frac{c}{3}\ln\left[\frac{L}{\pi}\sin\left(\frac{\pi L_A}{L}\right)\right].
\eeq
Here $L\ (L_A)$ is the system (partition) length, respectively. Therefore, the above measurement unambiguously indicates the microscopic realization of the $m\!=\!6$ SMM associated to the criticality of the lattice Hamiltonian (\ref{hami}).

\begin{figure}[tb]
\centering
\includegraphics[width=0.47\textwidth]{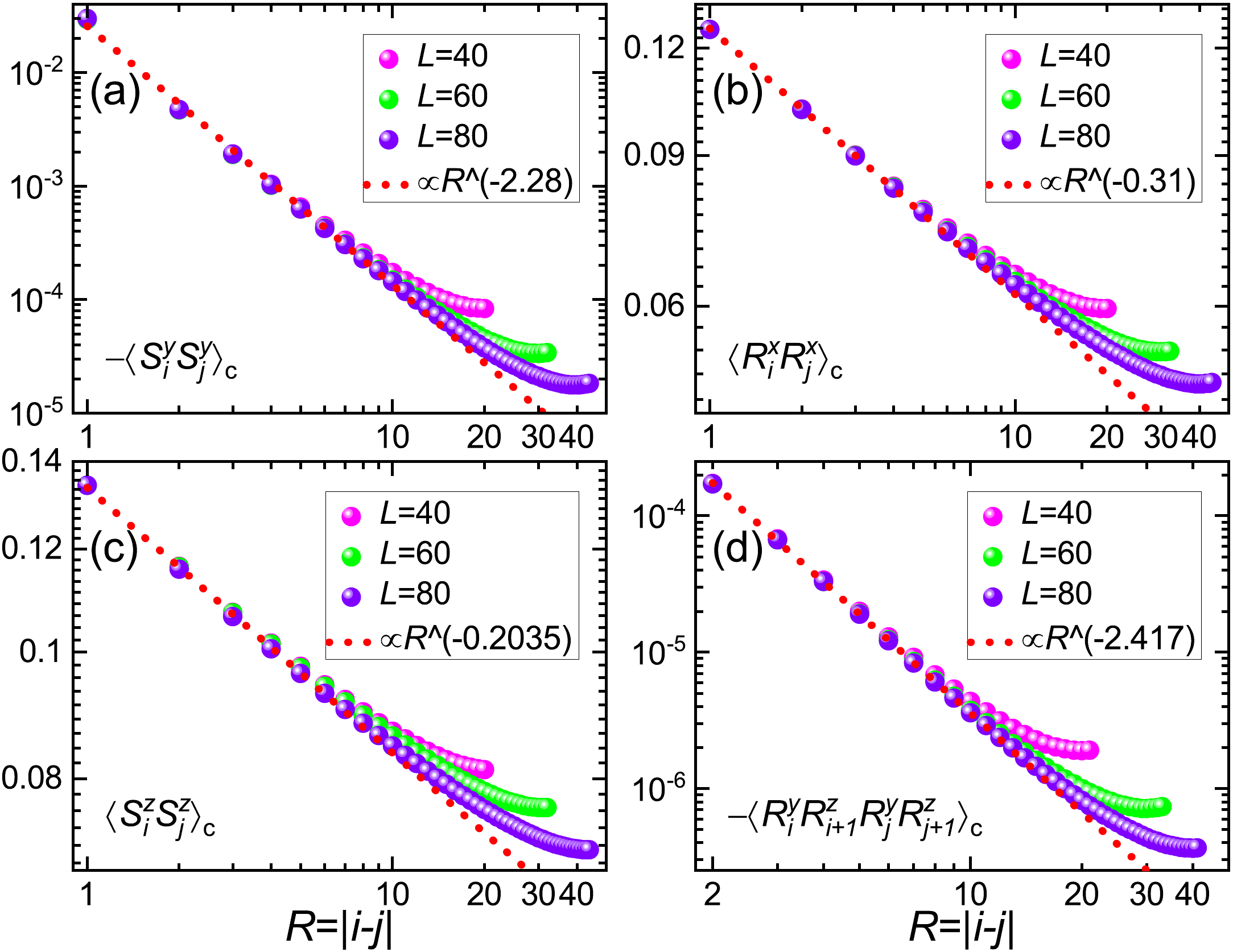}
\caption{\label{fig:fig2} Spatial profiles for various spin correlation functions calculated at the critical point: $g/J_{zz}\!=\!0.5,\ J_{xx}/J_{zz}\!=\!1.504$. The scaling dimension is measured from the exponent of the decay of the connected correlation function. The first column targets the $\widehat{\boldsymbol{\sigma}}$-spins, where the estimates of (a) and (c) are compatible with the Ramond HWs $9/16$ and $5/96$ in Kac table \ref{tab:1}, respectively. Here $\widehat{\boldsymbol{S}}_i\!\coloneqq\!\widehat{\boldsymbol{\sigma}}_i/2$ and $\widehat{\boldsymbol{R}}_i\!\coloneqq\!\widehat{\boldsymbol{u}}_{i,a(b)}/2$. Analogously, the right column demonstrates that $R^y_iR^z_{i+1}$ is the superpartner of $R^x_i$; the combination of the two resembles the NS superfield with conformal dimension $1/12$. Note the role of the spin-reversal symmetry in distinguishing the scaling behaviours between $S^y_i$ and $R^y_iR^z_{i+1}$. }
\end{figure}

To calibrate this nontrivial identification, we further probe at the transition point the varied superconformal scaling fields, which together with the central charge define the underlying SCFT, through examining the connected correlation functions involving the local spins. The conformal dimensions or highest weights (HWs) of the related primary fields in the Ramond (R) and Neveu-Schwarz (NS) sectors of the $m\!=\!6$ SMM have been listed on Table \ref{tab:1}. Under the PBCs of the spin chain, the primary field $\phi_{h,\widebar{h}}$ typically possesses zero conformal spin, i.e., $h-\widebar{h}=0$, and the corresponding scaling dimension is simply $h+\widebar{h}=2h$, so in the continuum limit, the two-point correlation function of the scaling field follows a generic power-law decay, $\langle \phi_{h,\widebar{h}}(z,\widebar{z})\phi_{h,\widebar{h}}(0,0)\rangle_{\textrm{c}}\!\approx\!1/|z|^{4h}$. In addition, one can also exploit the symmetry and duality of the model in helping clarify the assignment and organization of the selected primary fields in terms of the lattice spin operators. For the case at hand, the relevant unitary symmetry of model (\ref{hami}) turns out to be the $\mathbb{Z}_2$ spin-reversal transformation ${\sf R}_\sigma\!\coloneqq\!\prod_i\sigma^x_i$, which leaves $\widehat{\boldsymbol{u}}_{i,a(b)}$-spins and $\sigma^{x}_i$ intact, but flips $\sigma^{y,z}_i\!\rightarrow\!-\sigma^{y,z}_i$. (In Majorana representation of the $\widehat{\boldsymbol{\sigma}}$-spins, ${\sf R}_\sigma$ amounts to the fermion-parity operator.) It is to be expected that the conformal fields in the NS (R) sector should be even (odd) under ${\sf R}_\sigma$ by recalling that the JW transformation transmutes the spin-chain PBCs to the fermion-chain PBCs (Ramond) if the fermion parity is odd or to the fermion-chain anti-PBCs (Neveu-Schwarz) if the fermion parity is even. The calculated correlation functions in Fig.~\ref{fig:fig2} are in accord with this symmetry consideration. Particularly, from panels (a) and (c) the extracted conformal dimensions of $\sigma^y_i$ and $\sigma^z_i$ match the Ramond HWs $9/16$ and $5/96$ in Table \ref{tab:1}, respectively, suggesting the coarse-grained version of them assumes the role of the corresponding R spin fields. 

Furthermore, Hamiltonian (\ref{hami}) remains invariant under a duality transformation ${\sf D}$ \cite{LASSIG1991}. Similar to the GSV model, ${\sf D}$ incorporates the Kramers-Wannier duality of the $\widehat{\boldsymbol{\sigma}}$-spins with the time reversal and translation of the $\widehat{\boldsymbol{u}}$-spins: $\sigma^x_i\!\rightleftharpoons\!\sigma^z_i\sigma^z_{i+1},\ \widehat{\boldsymbol{u}}_{i,a}\!\rightarrow\!-\widehat{\boldsymbol{u}}_{i,b},\ \widehat{\boldsymbol{u}}_{i,b}\!\rightarrow\!-\widehat{\boldsymbol{u}}_{i+1,a}$. In the superspace formalism of the SCFTs, each NS field generically has a superpartner whose conformal dimension has been shifted by $1/2$ \cite{QIU1986}. The pair forms one superfield of the so-called spin model where the fermionic components of the superfield have been projected out \cite{FQS}. The NS field and its superpartner can be further differentiated upon implementing the duality transformation: the NS field is odd whereas the superpartner field is even under ${\sf D}$. In Figs.~\ref{fig:fig2}(b) and (d), we show that $(u^x_{i,\alpha},u^y_{i,\alpha}u^z_{i+1,\alpha})$ with $\alpha\!=\!a,b$ comprises such a NS superfield of conformal dimension $(1/12,7/12)$. Notice that both $u^x_{i,\alpha}$ and $u^y_{i,\alpha}u^z_{i+1,\alpha}$ are $\mathbb{Z}_2$-even operators of the spin-reversal symmetry ${\sf R}_\sigma$; however $u^x_{i,\alpha}$ becomes odd, while $u^y_{i,\alpha}u^z_{i+1,\alpha}$ stays even under the ${\sf D}$ transformation. Hence, in addition to the detection of the characteristic central charge $5/4$, identifying the lattice representations of the pair of NS fields as well as the selected R spin fields provides stronger evidence supporting the stabilization of the $m\!=\!6$ SMM at the critical point.

{\it Bosonization.---}To gain insights on the emergent superconformal criticality, a continuum field-theoretical description has been developed for $H$. As already mentioned, the self-dual TFI Hamiltonian $H_1$ can be fermionized upon a JW transformation, $\psi_{1}(j)\!\coloneqq\!\sigma^z_j\prod_{i<j}\sigma^x_i,\ \psi_{2}(j)\!\coloneqq\!\sigma^y_j\prod_{i<j}\sigma^x_i$. In terms of the chiral Majorana fields, $\chi_{R/L}\!\coloneqq\!\frac{1}{\sqrt{2}}\left(\psi_{1}\pm\psi_{2}\right)$, the Euclidean action of $H_1$ attains a massless form, $S_1/\hbar\!=\!\int d\tau dx\frac{1}{2}\widebar{\boldsymbol{\chi}}(\tau,x)\slashed{\partial}\boldsymbol{\chi}(\tau,x)$. Here by definition, $\slashed{\partial}\!\coloneqq\!\gamma^\mu\partial_\mu$, the spinor $\boldsymbol{\chi}^{\sf T}\!\coloneqq\!\left(\chi_R\ \ \chi_L\right)$, and its conjugate $\widebar{\boldsymbol{\chi}}\!\coloneqq\!\boldsymbol{\chi}^{\sf T}\gamma^0$. One convenient representation of the $2$D $\gamma$ matrices with the Euclidean signature, $\gamma^0\!=\!\sigma_2,\ \gamma^1\!=\!-\sigma_1$, has been chosen.

\begin{figure*}[t]
\centering
\includegraphics[width=1\textwidth]{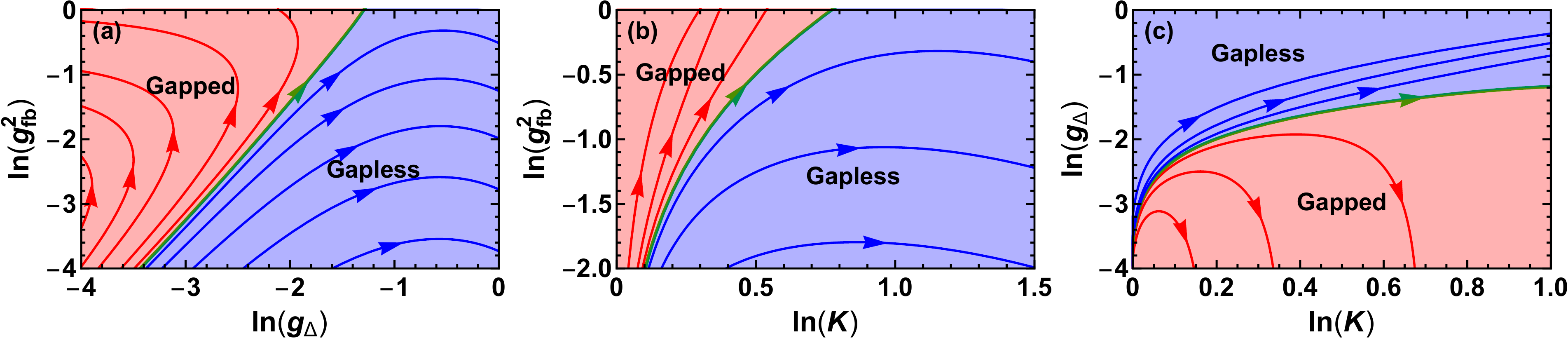}
\caption{\label{fig:fig3} The characteristic RG flows derived from action (\ref{action}) for the initial parameters $\Delta\!>\!0$ and $K\!\approx\!1$. These trajectories belong to a different category than the KT-type transition in the SSG model (accessible by $\Delta\!<\!0$). The DMRG results in Fig.~\ref{fig:fig1} demonstrate that along the critical line (the green curve) separating the gapped and gapless phases, the $m\!=\!6$ SMM arises. (a) displays the parametric dependence of $g_\Delta(\ell)$ and $g_\fb(\ell)$ in a log-log plot. The $2$D projections of the $3$D flows onto the $K$-$g_\fb$ and $K$-$g_\Delta$ planes are shown by (b) and (c), respectively.}
\end{figure*}

The strategy of handling the $\XZ$ part $H_2$ is to first fermionize the spin chain to obtain a quadratic Dirac-fermion model, and then apply the Abelian bosonization \cite{Giamarchi}. This procedure however is inapplicable to the case of TFI magnetic fluctuation where due to the fact that the underlying degrees of freedom are Majorana fermions, one should instead employ the Landau-Ginzburg-Wilson (LGW) $\phi^4$ theory. Accordingly, the $\XZ$ spin chain is first recast into a spinless free-fermion form, $H_2\!=\!\sum_j(-tc^\dagger_jc_{j+1}+\Delta c^\dagger_jc^\dagger_{j+1}+\textrm{H.c.})$, with $t\!\coloneqq\!J_{xx}+J_{zz},\ \Delta\!\coloneqq\!J_{xx}-J_{zz}$. Note that the fermion density is precisely $\langle c^\dagger_jc_j\rangle\!=\!1/2$, so the Fermi level has been pinned to $k_F\!=\!k_R\!=\!-k_L\!=\!\pi/(2a)$. After linearizing the band around the vicinity of the two Fermi points, the bosonization of the $c$-fermions in the right ($R$) and left ($L$) branches is subsequently achieved via the identity, $c_{R/L}(x)\!\rightarrow\!\frac{\eta_{R/L}}{\sqrt{2\pi\alpha}}e^{\pm ik_Fx}e^{i(\pm\theta+\phi)}$, where $[\theta(x),\phi(x')]\!=\!i\frac{\pi}{2}\textrm{sgn}(x-x')$ and the short-distance cutoff is about the lattice spacing, i.e., $\alpha\!\sim\!a$. Therefore, the bosonized action of $H_2$ takes a standard sine-Gordon expression, $S_2/\hbar=\frac{1}{2\pi}\int d\tau dx\!\left[\frac{1}{v}\!\left(\partial_\tau\phi\right)^2+v\!\left(\partial_x\phi\right)^2+\frac{4\Delta}{\hbar\alpha}\cos\!\left(2\phi\right)\right]$, where the boson velocity $v\!=\!2at/\hbar$.

We now turn to the coupling term $H_3$. The bosonization of the $\widehat{\boldsymbol{u}}$-spins is proceeded as usual \cite{Giamarchi} by introducing the associated ladder operators, $u^\pm_j\!\coloneqq\!u^z_j\pm i u^x_j$, where the relabeling convention stated above Eq.~(\ref{dualspin}) has been used. As the JW string possesses a local form in terms of the nonchiral bosonic fields $\theta$ and $\phi$, in the continuum limit, the raising operator can be expressed as $u^+_j\!\rightarrow\!\sqrt{\frac{2a}{\pi\alpha}}\left[e^{-i\phi}+e^{-i\phi}(-1)^{x/a}\cos(2\theta)\right]$. The fermionization of the $\widehat{\boldsymbol{\sigma}}$-spins is straightforward. Rewriting the JW transformation as $i\psi_{2}(j)\psi_{1}(j\!+\!1)\!=\!\sigma^z_j\sigma^z_{j+1},\ i\psi_{1}(j)\psi_{2}(j)\!=\!\sigma^x_j$ and combining it with the bosonized form of the $u^x$-spins lead to a boson-fermion coupled expression, $H_3\!\approx\!g\sqrt{\frac{8a}{\pi\alpha}}\int dx\left\{\cos[\phi(x)]\cdot\widebar{\boldsymbol{\chi}}(x)\boldsymbol{\chi}(x)\right\}$. Collecting together all pieces of derivation yields the desired action of $H$ $(\hbar=1)$,
\begin{align}
S&=\!\int\!d\tau dx\!\left\{\frac{1}{2}\widebar{\boldsymbol{\chi}}\slashed{\partial}\boldsymbol{\chi}+\frac{K}{2\pi v}\!\left(\partial_\tau\phi\right)^2+\frac{Kv}{2\pi}\!\left(\partial_x\phi\right)^2 \right. \nonumber \\
&\left.\ \ \ \ \ \ \ \ \ \ \ \ \ \ \ \ \ \ \ \ +\frac{2\Delta}{\pi\alpha}\cos\!\left(2\phi\right)+g\sqrt{\frac{8a}{\pi\alpha}}\cos(\phi)\cdot\widebar{\boldsymbol{\chi}}\boldsymbol{\chi}\right\}\!, \label{action}
\end{align}
where the bare Luttinger parameter $K\!=\!1$. The Euclidean action (\ref{action}) resembles the well-known supersymmetric sine-Gordon (SSG) model \cite{WITTEN1978,Shankar,GoldschmidtNPB}, but this is not entirely true, because the SSG model normally flows to a massive phase in the infrared (IR) which contradicts our DMRG results.

{\it RG analysis.---}A closer inspection reveals that depending on the relative arrangements of the last two terms in (\ref{action}), there could exist two inequivalent theories described by $S$. For instance, shifting $\phi\!\rightarrow\!\phi+\pi$ shows that the sign of $g$ is inessential. Whereas, shifting $\phi\!\rightarrow\!\phi+\frac{\pi}{2}$ suggests that the switch of the sign of $\Delta$ has to be accompanied by a simultaneous change of the $g$-term from $\cos(\phi)$ to $\sin(\phi)$. Therefore, without loss of generality, if keep staying with both cosine functions of $\phi$ in (\ref{action}), $\Delta>0$ and $\Delta<0$ will give rise to two different types of theories that cannot be connected by simply varying the field $\phi$. This bifurcation structure apparently does not occur in the SG model. 

The above heuristic argument can be formulated quantitatively under the scheme of perturbative RG. Following Cardy \cite{CardyBook}, we implement the standard RG procedure in real space by utilizing the technique of operator product expansion (OPE) applicable at the critical region with the emergent Lorentz invariance assumed. At one-loop order, the set of RG flow equations is derived as follows, 
\begin{align}
\frac{dK(\ell)}{d\ell}&=\pi^2 g^2_{\Delta}(\ell)+\frac{1}{2}g^2_{\fb}(\ell), \label{rg1} \\[0.5em]
\frac{dg_\Delta(\ell)}{d\ell}&=\left[2-\frac{1}{K(\ell)}\right]g_{\Delta}(\ell)-\frac{1}{\pi}g^2_{\fb}(\ell), \label{rg2} \\[0.5em]
\frac{dg_{\fb}(\ell)}{d\ell}&=\left[1-\frac{1}{4K(\ell)}\right]g_{\fb}(\ell)-\frac{\pi}{2}g_{\Delta}(\ell)g_{\fb}(\ell). \label{rg3}
\end{align}
Here the dimensionless coupling constants $g_\Delta\!\propto\!2\Delta a/(\pi v)$, $g_\fb\!\propto\!2\sqrt{2}g a/(\sqrt{\pi} v)$, and $K\!\sim\!1$. One salient feature of Eqs.~(\ref{rg1})--(\ref{rg3}) is that if initially $g_\Delta\!<\!0$, then both $g_\Delta$ and $g_\fb$ will grow monotonically along the RG trajectory. This growth in turn contributes to the increase of $K$ as well. Consequentially, the whole system displays the typical behaviour of the so-called supersymmetric Kosterlitz-Thouless (KT) transition \cite{Goldschmidt,GoldschmidtNPB}, and eventually resides in a gapped massive phase. This result is consistent with the common expectation, i.e., if $\Delta\!<\!0$ in Eq.~(\ref{action}), the system will generally flows toward the SSG model in the IR with emergent SUSY but zero central charge \cite{WITTEN1978}. Note, in addition, that the sign of $g_\fb$ in (\ref{rg1})--(\ref{rg3}) can be tuned at will, a reflection of the underlying time-reversal symmetry (i.e., the $\pi$-phase shift of $\phi$) in (\ref{action}).

More interestingly, the landscape of the RG trajectory changes qualitatively if $g_\Delta$ instead starts from the appropriate domain of the positive values. [Recall that in the numerical simulation of the lattice model (\ref{hami}), the critical $J^c_{xx}$ in Fig.~\ref{fig:fig1}(b) corresponds to $\Delta\!=\!0.504$.] In this circumstance, rather than mutual facilitation, the two channels $g_\Delta$ and $g_\fb$ turn to compete with each other. Their compromise hence necessarily dictates the existence of a critical region where the probable exotic properties can emerge. Figure \ref{fig:fig3}(a) is the \lq\lq phase diagram'' of (\ref{action}) with $\Delta\!>\!0$ obtained from solving the RG Eqs.~(\ref{rg1})--(\ref{rg3}) collectively. In accord with the DMRG results presented in Fig.~\ref{fig:fig1}, once $g_\fb$ prevails over $g_\Delta$, the relevant $g$-term in (\ref{action}) serves the dual role of the mass to the Majorana fermion $\boldsymbol{\chi}$ and the SG term of the real boson $\phi$, thus leading to a fully gapped phase. In comparison, if $g_\fb$ fades away, the chain system gets effectively decoupled; although the real boson is gapped out by the dominating $\Delta$-term, the critical real-fermion mode survives and contributes to the single gapless channel of central charge $1/2$. Thereby, the transition line where competing strengths $g_\fb$ and $g_\Delta$ reach a balance locates right at the interface of these two distinct phases, as is depicted by the green curve in Fig.~\ref{fig:fig3}(a). Overall, the qualitative agreement between the DMRG phase diagram [Fig.~\ref{fig:fig1}(a)] and the analytical phase diagram [Fig.~\ref{fig:fig3}(a)] is telling. To our knowledge, this second type of theory in (\ref{action}) as well as the associated RG flows in Fig.~\ref{fig:fig3} have not been fully studied before, in particular, their close relation to the $m\!=\!6$ SMM.

\def\arraystretch{1.25}
\setlength{\tabcolsep}{18pt}
\begin{table}[b] 
\caption{Compare the IR and UV fixed points of the two theories captured by action (\ref{action}). See main text for the explanation.} 
\label{table2} 
\centering
\begin{tabular}{ ccc }
 \hline\hline
 \\[-0.9em]
 & IR & UV \\ [0.5em] \cline{1-3} \\ [-0.9em]
 $\Delta\!<\!0$ & massive SUSY & $\mathcal{W}\!=\!\lambda\cos(\beta\Phi)$  \\ [0.5em] \cline{2-3} \\ [-0.9em]
 $\Delta\!>\!0$ & $m\!=\!6$ SMM & $\mathcal{W}\!=\!i\lambda\sin(\beta\Phi)$ \\[0.3em]
 \hline\hline
\end{tabular}
\end{table}

{\it Superspace and Feigin-Fuchs construction.---}If action (\ref{action}) with positive $\Delta$ is not manifestly supersymmetric, where does the IR superconformal symmetry stem from? To set the stage of tackling this question, let's first recall the superspace formalism \cite{Shifman} of the SSG model. Promote the $2$D Euclidean spacetime to a superspace by adding the Grassmann variable $\boldsymbol{\theta}^{\sf T}\!\coloneqq\!\left(\theta_0\ \ \theta_1\right)$, then the normalized real superfield assumes $\Phi\!\coloneqq\!\frac{1}{\sqrt{4\pi}}\phi+\frac{1}{\sqrt{2\pi}}\widebar{\boldsymbol{\theta}}\boldsymbol{\chi}+\frac{1}{2}\widebar{\boldsymbol{\theta}}\boldsymbol{\theta}F$, where $\widebar{\boldsymbol{\theta}}\!\coloneqq\!\boldsymbol{\theta}^{\sf T}\gamma^0$ and $F$ an auxiliary field. Adopt this notation, the minimal $2$D Wess-Zumino model \cite{WESS1974} can be written as $S_{\textrm{WZ}}=\int d^2\theta d\tau dx\left[\frac{1}{2}\widebar{D}_\alpha \Phi D_\alpha\Phi+2 \mathcal{W}(\Phi)\right]$ with spinorial derivative $D_\alpha\!\coloneqq\!\partial_{\widebar{\boldsymbol{\theta}}_\alpha}+(\gamma^\mu\boldsymbol{\theta})_\alpha\partial_\mu$ and $\int d^2\theta\widebar{\boldsymbol{\theta}}\boldsymbol{\theta}\!=\!-1$, which is invariant for any superpotential $\mathcal{W}$ under the SUSY transformations, $\delta\phi=\widebar{\boldsymbol{\varepsilon}}\boldsymbol{\chi},\ \delta\boldsymbol{\chi}=\partial_\mu\phi\gamma^\mu\boldsymbol{\varepsilon}+F\boldsymbol{\varepsilon},\ \delta F=\widebar{\boldsymbol{\varepsilon}}\gamma^\mu\partial_\mu\boldsymbol{\chi}$. Here $\boldsymbol{\varepsilon}$ stands for an infinitesimal change along the Grassmann coordinate. Supply $S_{\textrm{WZ}}$ with $\mathcal{W}\!=\!\lambda\cos(\beta\Phi)$ yields the SSG action.

We have revealed that action (\ref{action}) allows two categories of theories distinguished by the sign of $\Delta$. As summarized in Table \ref{table2}, the first type is of $\Delta\!<\!0$ and amounts to the SSG model in the superspace formalism whose IR and UV behaviours are well studied. While, for the second type of theory featuring $\Delta\!>\!0$, our combined DMRG and bosonization analyses have provided corroborating proof demonstrating the stabilization of the $m\!=\!6$ SMM at its IR fixed point. Then, what would be the potential candidate for the corresponding UV theory? From examining the interrelationships among the four entries of Table \ref{table2}, it is tempting to conjecture that the UV fixed point of $\Delta\!>\!0$ might be effectively described by an unconventional SSG model whose superpotential is purely imaginary \cite{FendleySaleurZamolodchikov}. Indeed, in parallel to the distinction between the situations of \emph{negative} and \emph{positive} $\Delta$, it is known from the conformal perturbation theory that upon a Feigin-Fuchs construction (or Coulomb-gas formalism) \cite{DotsenkoFateev,BigYellow,BERNARD1990,Ahn1990,AHN1991} the \emph{real} and \emph{imaginary} SG superpotentials will lead respectively to the massive and massless phases in the IR limit. Specifically, feed $\mathcal{W}\!=\!i\lambda\sin(\beta\Phi)$ into $S_{\textrm{WZ}}$ yields
\beq
\mathcal{L}_{\textrm{WZ}}=\left(\frac{1}{2}\widebar{D}_\alpha \Phi D_\alpha\Phi-\lambda \mathcal{V}_-\right)+\lambda \Phi_{1,3},
\eeq
where a background charge $\alpha_0\!=\!1/\sqrt{2m(m+2)}$ has been implicitly injected such that when $\beta^2\!=\!\frac{8\pi m}{m+2}$, the $e^{-i\beta\Phi}$ piece in $\mathcal{W}$ becomes the screening operator $\mathcal{V}_-$ of conformal dimension $\frac{1}{2}$ (remember the real-fermion field $\boldsymbol{\chi}$ will contribute the other $\frac{1}{2}$ \cite{BKT1985}), which combined with the free Lagrangian describes the $\mathcal{N}\!=\!1$ SMM at UV with central charge $c_{\textrm{UV}}(m)\!=\!\frac{3}{2}-24\alpha^2_0$. The remaining vertex operator $e^{i\beta\Phi}$ now has dimension $\frac{m-2}{2(m+2)}$, thus being equivalent to the primary field $\Phi_{1,3}$ which, acting as the quasi-marginal deformation, can drive the \emph{massless} RG flows \cite{KASTOR1989,KITAZAWA1988} from $c_{\textrm{UV}}(m)\!\rightarrow\! c_{\textrm{IR}}(m-2)$. Accordingly, in the case of $\Delta\!>\!0$, action (\ref{action}) may sit in the attractive basin of the UV fixed point of $\mathcal{L}_{\textrm{WZ}}$ so that the two theories can be linked via an analytic continuation and share the same IR fixed point with the superconformal symmetry ensued.

{\it Supersymmetry breaking.---}In addition to the emergent SUSY, Fig.~\ref{fig:fig1}(b) also demonstrates its spontaneous breaking when tuned away from the criticality. The generated massless Majorana fermion can be interpreted as the resulting Goldstino from this perspective. However, Fig.~\ref{fig:fig2}(c) simultaneously illustrates that the self-symmetric Ramond HW $c/24$ is present at the superconformal point as well. As per the common wisdom, for even-$m$ SSMs, the SUSY is unbroken iff there is a ground state in the sector of $h_\R\!=\!c/24$ \cite{FQS,KASTOR1989}. This seemingly contradictory observation actually lends support to a recent work by the authors \cite{ChenMaciejko} where it has been proved that the modular invariance of the torus partition function necessitates the symmetrization of the Ramond ground-state manifold associated to $h_\R\!=\!c/24$, hence permitting the breaking of SUSY in this circumstance and the vanishing of the Witten index. 

{\it Conclusion.---}We have put forward a lattice Hamiltonian whose criticality belongs to the $m\!=\!6$ superconformal universality class. The extended DMRG and continuum calculations have suggested that unlike the previous proposals, most of which are based on the TCI model and the supersymmetric LG theory, the emergent SUSY in this new kind of model is mainly derived from a generalized SSG-like action and its massless RG flows. Through introducing such a \lq\lq vehicle'' that might also be feasible in experiments, our work opens the prospects for the future studies on the even-$m$ series of SMMs, which will deepen and refine our understanding on the global structure of SUSY and help facilitate its realization in low-dimensional quantum many-body systems.

We are grateful to Prof.~Joseph Maciejko for the useful discussion and Prof.~Suk Bum Chung for the initial collaboration on the related topic and for the access to the computational resource. We also acknowledge the support from the University of Alberta.

\bibliography{SSG}

\end{document}